\def \JMP {{\it J. Math. Phys. }}

\def \bc {\begin{center}}
\def \ec {\end{center}}

\def \bfr {\begin{flushright}}
\def \efr {\end{flushright}}

\def \v {\vskip}
\def \ii {\'\i}

\def \ba {\begin{array}}
\def \ea {\end{array}}

\def \bea {\begin{eqnarray}}
\def \eea {\end{eqnarray}}

\def \be {\begin{equation}}
\def \ee {\end{equation}}

\def \L {{\cal L}}%% Usually the lagrangian
\def \d {\hbox{d}\,}%% The exterior derivative operator

\def \p {\partial}

\def \H {{\cal H}}
\def \x {{\bf x}}

\def \dirac {{\slash\!\!\!}}
\documentstyle[12pt,a4]{article}

\begin{document}
\def \thesubsection {\Alph{subsection}}
\def \thesubsubsection {\thesubsection.\arabic{subsubsection}}

%%%%%%%%%%%%%%%%%%%%%%%%%%%%%%%%%%%%%%%%%%%%%%%%%%%%%%%%%

%%%%%%%%%%%%%%%%%%%%%%%%%%%%%%%%%%%%%%%%%%%%%%%%%%%%%%
%%%%%%%%%%%%%%%%%%%%%PRINCIPIO DEL DOCUMENTO%%%%%%%%%%%%%%%%
%%%%%%%%%%%%%%%%%%%%%%%%%%%%%%%%%%%%%%%%%%%%%%%%%%%%%%%%%%%
\thispagestyle{empty}
%\noindent{\bf Preliminary Version}

%\today

%\noindent hep-th/9805010
% Password:  d7v3k

\begin{center}

{\bf QUANTUM AND CLASSICAL FIELDS IN THE FINITE-DIMENSIONAL
FORMALISM}

{\it Miguel Navarro}\footnote{e-mail: nim314159@aol.com}
\end{center}
\v4mm

\noindent Instituto Carlos I de F\ii sica Te\'orica y
Computacional. Granada, Spain \footnote{On leave of absence.
%Please all correspondence to 101, Iffley Road, London W6 0PD.
%United Kingdom.
}.

\v5mm
\centerline{\bf Abstract}
\v2mm
\footnotesize

The quantization rules recently proposed by M. Navarro
\cite{[towardfqft]} (and independently I.V. Kanatchikov
\cite{[Kanatchikov]}) for a finite-dimensional formulation of
quantum field theory are applied to the Klein-Gordon and the Dirac
fields to obtain the quantum equations of motion of both fields.
In doing so several problems arise.

 Solving these difficulties
leads us to propose a new classical canonical formalism, which, in
turn, leads us to new, improved rules of quantization. We show
that the new classical equations of motion and rules of
quantization overcome several known unsatisfactory features of the
previous formalism. We argue that the new formalism is a general
improvement with respect to the previous one.

Further we show that the quantum field theory of the Dirac and
Klein-Gordon field describes particles with extra,
harmonic-oscillator-like degrees of freedom. We argue that these
degrees of freedom should give rise to a multi-particle
interpretation of the formalism.

 \v3mm \noindent PACS numbers: 03.65.Bz, \ 11.10.Ef,
\ 03.70.+k\hfil\break \noindent Keywords: Quantum Mechanics, Field
Theory, Equations of Motion, Canonical Formalism. \normalsize
\newpage
\setcounter{page}{1}
\section{Introduction}

At present the main goal of Theoretical Physics is to produce an
unified quantum theory of all forces of nature, including gravity.
Unfortunately, and despite the tremendous success of both these
theoretical frameworks, neither General Relativity nor, above all,
Quantum Field Theory are theories with firm grounds that would
provide confidence for further building. Instead, it seems that
both theories will have to be dramatically reformulated before a
true unification can be devised.

An approach that may help in this regard is the finite-dimensional
formulation of field theory. This approach has a long history,
which dates back to the thirties \cite{[antecendentes]}. Much of
the work done on the formalism focused on the classical theory or
on following routes to the quantum theory which closely mimic the
one which, starting from Classical Mechanics, leads to the
standard Quantum Mechanics.

Recently, however, a change of strategy has taken place and the
focus has shifted towards formulating a quantum theory without
paying much attention to 'deriving' it from the classical
formalism. In particular, inspired by a previous proposal by Good
\cite{[Good]} rules of quantization and equations of motion has
been postulated by M. Navarro \cite{[towardfqft]} (and
independently by I.V. Kanatchikov \cite{[Kanatchikov]}) that give
rise to a promising finite-dimensional formulation of QFT. In this
paper that formalism is applied to the scalar field and the Dirac
field. In doing so several unsatisfactory aspects of the classical
and quantum formalisms become apparent. In order to overcome these
problems a modification of the classical as well as the quantum
formalism is proposed.

The paper is organised as follows. In section 2 we review the
finite-dimensional canonical formalism for the classical fields.
In section 3 the formalism put forward by Navarro and Kanatchikov
is reviewed. In section 4 we discuss the interpretation of the
formalism. In section 5 the formalism is applied to the scalar
field and the Dirac field to obtain the quantum equations of
motion. In section 6 we present the improved classical canonical
formalism. In section 7 improved rules of quantization are
presented to be applied to the Dirac and the scalar field. In
section 9 we state our conclusions and related comments.

\section{The finite-dimensional formulation of the classical field theory}

Currently we do not have a quantum theory but classical theories
and rules of quantization. The standard way of quantizing a field
theory relies on the fact that Classical Field Theory (CFT) can be
regarded as a generalization of Classical Mechanics (CM) in which
the finite number of degrees of freedom of the latter is replaced
with an infinite (continuum) number in the former. In this
formulation the fields are considered to be functions
$\varphi^a(\x)\,(t)\equiv\varphi^a_{\x}(t)$; that is, the spatial
co-ordinates are regarded as labels (the discrete superindex $a$
labels the different fields in the theory). This description is
supported primarily by the fact that it is a direct generalization
of Quantum Mechanics (QM), which, as a theory with a vast range of
successful predictions, is a source of great confidence. The
standard framework requires, nonetheless, the use of functionals
as well as infinite-dimensional differential calculus, which is
plagued with ambiguities. These ambiguities are at the root of the
renormalization problem.

There is however, a different way of looking at CFT as a
generalization of Classical Mechanics \cite{[antecendentes]}
\cite{[commentsGood]}. In this reading of CFT, all the
co-ordinates of the space-time are considered to be on the same
footing, no special role is played by time. The fields are not
taken to be an infinite (continuum) set of functions of time but
rather a discrete set of functions of all the space-time
co-ordinates: $\varphi^a=\varphi^a(x),$ with $x=(\x, t)$ and $a$ a
discrete label. Since there is a finite number of functions we
refer to this approach as the finite-dimensional formulation of
field theory as opposed to the standard or infinite-dimensional
formalism.

In this (finite-dimensional) formulation of field theory the
canonical formalism is normally assumed to be as follows. Given a
Lagrangian $\L = \L(\phi^a,\p_\mu\varphi^a)$, the covariant
momenta $\pi^\mu_a$ are defined by:

\be \pi^\mu_a= \frac{\p\>\L}{\p\>(\p_\mu\varphi^a)}\label{a2}\ee
Then the covariant Hamiltonian $\H$ is obtained by means of the
generalized covariant Legendre transform:

\be \H = \pi_a^\mu\p_\mu\varphi^a -\L\>.\label{a1b}\ee

If we now write the Lagrangian in the following covariant
Hamiltonian form

\be \L = \pi_a^\mu\p_\mu\varphi^a -
\H(\varphi^a,\pi_a^\mu)\label{a3}\ee its Lagrange equations of
motion will also have a covariant Hamiltonian form:

\bea \p_\mu\varphi^a &=& \frac{\p\>\H}{\p\>\pi^\mu_a}\\
\p_\mu\pi^\mu_a&=& -\frac{\p\>\H}{\p\>\varphi^a}\label{a4}\eea

In this context a mechanical system correspond to a field theory
in $1+0$ dimensions, i.e. in a spacetime with $1$ temporal and $0$
spacelike dimensions. When the system described is mechanical, the
finite-dimensional covariant formalism reduces to the ordinary one
(which we present here for later use).

 In the ordinary
canonical formalism the momenta associated to the Lagrangian $L=
L(\dot q^i, q^j)$ are defined by

\be p_i = \frac{\p\>L}{\p\>\dot q^i}\label{mechanicalmomenta}\ee
and the Hamiltonian is defined by

\be H(p_i, q^j) = p_i \dot q^i - L\label{mechanicalhamiltonian}\ee
which gives rise to the equation of motion

\bea \dot q^i &=& \frac{\p H}{\p\>p_i}\\
\dot p_i&=& -\frac{\p H}{\p\>q^i}\label{mechanicaleofm}\eea

\section{The finite dimensional formulation of quantum field theory}

The challenge is to translate the elegant, finite-dimensional
formulation of classical field theory briefly reviewed above to
the quantum theory. For doing so suitable rules of quantization
and equations of motion are needed.

Let us review the case of ordinary Quantum Mechanics, i.e. the
rules of quantization for a mechanical system.

The rules of quantization are
\bea q^i&\longrightarrow& \widehat{q}^i=q^i\nonumber\\
p_i&\longrightarrow& \widehat q_i= -i\frac{\p}{\p q^i}
\label{ordinaryqrules}\eea

The quantum theory is described by a (wave) function $\Psi(q^i,
t)$, which obeys the following equation of motion (the
Schr\"odinger equation):

\be i\frac{\d}{\d t}\Psi = \widehat{H}\Psi\label{Schrodinger}\ee
The goal is to generalize the formalism described by Eq.
(\ref{ordinaryqrules}-\ref{Schrodinger}) to the general case
described by Eq. (\ref{a2}-\ref{a4})

Inspired by previous work by Good \cite{[Good]} (see also
\cite{[commentsGood]}) a proposal was recently put forward by M.
Navarro \cite{[towardfqft]} (and independently by I.V. Kanatchikov
\cite{[Kanatchikov]}) for quantization rules and evolution
equations in the finite-dimensional formalism.

To motivate the proposal, let us consider the ordinary harmonic
oscillator and the Dirac field. The respective Lagrangians can be
written:

\bea  \L_{HO} &=& a^*(i\dot{a} - a)\label{c1}\\
\L_D &=& \bar{\varphi}(i\dirac{\p}\varphi-\varphi)\label{c2}\eea
where $a$ ($a^*$) is the annihilation (creation) operator,
$\dirac{\p}\equiv \gamma^\mu\p_\mu$, with $\gamma_\mu$ the Dirac's
matrices, and $\bar{\varphi}=\varphi^\dagger\gamma^0$.

Eqs. (\ref{c1}) and (\ref{c2}) tell us
that the Dirac field is a higher-dimensional
generalization of the ordinary harmonic oscillator. The generalization
is accomplished by replacing the time derivative ${\d}/{\d t}$
with the operator $\dirac{\p} =\gamma^\mu\p_\mu$.

Mimicking that generalization,  the following equation of motion
for our finite-dimensional QFT was postulated
\cite{[towardfqft]}\cite{[Kanatchikov]}.

\v2mm \noindent{\bf Quantum equation of motion}

\be i\dirac{\p}\Psi = \widehat{\H}\Psi, \qquad
\dirac{\p}\equiv\gamma^\mu \p_\mu \label{neweqofmot}\ee Here
$\gamma_\mu$ are quantities which play a role similar to Dirac's
matrices in the relativistic theory of the electron. In the
present letter, and for the sake of clarity, we will identify
these quantities with the Dirac matrices. The developments in the
present paper suppport this identification but we should keep in
mind, nevertheless, that further developments of the theory may
require that identification to be dropped.

The next step is to construct the operator $\widehat{\H}$. That
is, we need quantization rules. The proposal in Ref.
\cite{[towardfqft]}\cite{[Kanatchikov]} is the following:

\v2mm

\noindent{\bf Quantization rules}

\bea \varphi^a&\longrightarrow& \widehat{\varphi}^a=\varphi^a\nonumber\\
\pi^\mu_a&\longrightarrow& \widehat\pi^\mu_a=
-i\gamma^\mu\frac{\p}{\p\varphi^a} \label{newrules}\eea

The rules of quantization (\ref{newrules}) and the evolution
equation (\ref{neweqofmot}) fulfill the following properties.
\begin{enumerate}

\item Both rules of quantization and
equations of motion are explicitly covariant; i.e., space and time
co-ordinates are treated on the same footing.

\item Within the limits of mechanical systems these rules of
quantization and equations of motion reduce themselves to the
familiar canonical rules of quantization and Schr\"odinger
equation of evolution of ordinary Quantum Mechanics.

\item The equations of evolution are
second order in derivatives and first order in derivatives of the
space-time co-ordinates.
\end{enumerate}

Most notably, and unlike in Good's proposal, ordinary Quantum
Mechanics is contained in this new proposal. Therefore, the vast
amount of experimental predictions of ordinary QM is entirely and
automatically incorporated into our proposal. Hence, to rule out
the new proposal we would have to look for a test which implied a
genuine field system.

\section{Interpretation of the quantum theory}

The quantum theory is described by a wave function
$\Psi(\varphi,x)$. By analogy with mechanical systems the
temptation would be to interpret the wave function as the
probability density of finding a value $\varphi$ of the field in
the point $x$ of the spacetime. With this interpretation, the
probability of finding any value of $\varphi$ in the point $x$
would be unity, which would require

\be \int_{-\infty}^{+\infty}\d \varphi |\Psi(\varphi,x)|^2=1\ee

This interpretation was discussed in Ref. \cite{[towardfqft]}
where it was shown that it leads to serious difficulties. These
obstacles could perhaps be overcome in ways worth exploring. It
turns out however that there is a different and more natural
interpretation of the wave function $\Psi(u,x)$.

From eq. (\ref{neweqofmot}) it follows that the current

\be j^\mu=\int \d\varphi \bar{\Psi}\gamma^\mu\Psi \label{jmu}\ee
is conserved as long as the Hamiltonian $\hat{\cal H}$ is
self-adjoint with respect to $\int \d \varphi$ (which seems a
reasonable assumption to make).

This leads to an interpretation of the wave function according to
which

\be  P(\varphi,x)=\bar{\Psi}\gamma^0\Psi\label{probability}\ee
gives the probability density of finding an excitation $\varphi$
of the field in (the vicinity of) the point $x$. This
interpretation of the theory is in fact more appealing and
intuitively clear than the interpretation of the standard
formalism, which involves configurations of the field over
spacelike hypersurfaces.

Moreover, this interpretation clearly maintain the particle
interpretation of the theory - but with extra, 'internal' degree
of freedom.

\section{The quantum fields}

\subsection{The scalar field}

The Lagrangian for the scalar field is:

\be \L_S = \frac12(\p_\mu u\p^\mu u - m^2 u^2
)\label{LagrangianS}\ee

The momenta are given by ($u_\mu \equiv \p_\mu u$)

\be \pi^\mu = \frac{\p \L}{\p u_\mu} =  \p^\mu u \ee The classical
covariant Hamiltonian is given by

\be \H = \frac12(\pi_\mu \pi^\mu + m^2 u^2)\ee The rules of
quantization (\ref{newrules})
 yield the following operator

\be \widehat{\H} =  \frac12(-4\frac{\p^2}{\p u^2 } +m^2 u^2) \ee
and quantum equation of motion

\be i \dirac{\p} \Psi= \frac12(-4\frac{\p^2}{\p u^2 } +m^2
u^2)\Psi \label{eoms}\ee Eq. (\ref{eoms}) can be interpreted as
describing a Dirac field with a mass that is not constant but has
the dynamics of a harmonic-oscillator. Clearly it is expected that
the equally-spaced levels of this harmonic oscillator correspond
to different particle numbers in the system.

\subsection{The Dirac field}

The Lagrangian for the Dirac field is

\be \L_D = i\bar{v}\gamma^\mu \p_\mu v - m\bar{v}v
\label{LagrangianD}\ee

The canonical momenta are given by

\be \pi^\mu = i\bar{v}\gamma^\mu \label{momentaD}\ee This equality
means that the Lagrangian (\ref{LagrangianD}) is already written
in the canonical form (\ref{a3}) and that the covariant
Hamiltonian is

\be  \H = m\bar{v}v   \label{HamiltonianD}\ee Now, we can use the
rules of quantization (\ref{newrules}) to produce the operator

\be \widehat{\H} = m\frac{\p}{\p v}v\ee and the quantum equation
of motion

\be i \gamma^\mu \p_\mu \Psi=  m\frac{\p}{\p v}v\Psi
\label{eomD}\ee Clearly here there can be ordering issues that
should be taken into consideration when further developing the
theory.

Eq. (\ref{HamiltonianD}) closely mimics the mechanical harmonic
oscillator (\ref{c2}), which after all inspired our
finite-dimensional, covariant formalism.

\section{Improved classical formalism}

There is a serious problem, however, with the Dirac field as
discussed above: the covariant Hamiltonian (\ref{HamiltonianD})
together with the {\it classical} canonical equations of motion
(\ref{a4}) do not yield the correct {\it classical} equation of
motion (Dirac equation) for the Dirac field. Moreover, trying to
find a (different) covariant Hamiltonian by using the requirement
that it reproduces the correct classical equations of motions
leads to nowhere because the equations of motion (\ref{a4}) tend
to produce second-order equations of motion.

This difficulty highlights the fact that there are actually a
number of aspects of the classical canonical formalism
(\ref{a2}-\ref{a4}) that are unsatisfactory. Most notable among
them is the lack of a symmetry fields-momenta. This symmetry is
one of the principal features of the ordinary canonical formalism
and one of the reasons why the canonical formalism is used at all.
In fact, in the ordinary canonical formalism no distinction should
be made between fields and momenta since these quantities should
be interpreted as coordinates of the phase space and there is no
intrinsic reason to identify some of the coordinates of the phase
space with fields and the others with momenta. Moreover, this
symmetry is carried through towards ordinary quantum theory where
it plays an important role.

There is therefore a strong case for modifying the (classical)
formalism. Our proposal goes as follows.

For a Lagrangian $\L = \L(\varphi^a,\p_\mu\varphi^a)$ we define
the momenta

\be \pi_a =
\gamma_\mu\frac{\p\>\L}{\p\>(\p_\mu\varphi^a)}\label{newmomenta}\ee
The Hamiltonian $\H$ is defined by

\be \H(\varphi^a,\pi_a) = \pi_a\gamma^\mu\p_\mu\varphi^a
-\L\label{improvedcoH}\ee

If the Lagrangian is written in the form

\bea \L &=& \pi_a\gamma^\mu\p_\mu\varphi^a -
\H(\varphi^a,\pi_a)\nonumber\\
&=& \pi_a\dirac{\p}\varphi^a - \H(\varphi^a,\pi_a)
\label{improvedlagrangian}\eea it leads to the following classical
canonical equations of motion:

\bea \dirac{\p}\varphi^a &=& \frac{\p \H}{\p\>\pi_a}\nonumber\\
 \dirac{\p}\pi_a&=& -\frac{\p
H}{\p\>\varphi^a}\label{improvedofm}\eea

The new classical canonical formalism have several advantages when
compared with the old one. Most notably it is symmetric under the
change fields $\leftrightarrow$ momenta and it reproduced the
correct equations of motion for the Dirac field with the covariant
Hamiltonian $\H = m \bar{v}v$.

Aesthetically the new formalism is also more like the ordinary
formalism. Like in classical mechanics we have a single momenta
(albeit it is a matrix-valued one) for each field. The
generalization from one to the other is accomplished by replacing
everywhere the temporal derivative operator $\frac{\d}{\d t}$ with
the Dirac operator $\dirac{\p}$.

\section{Improved rules of quantization}

The new classical canonical formalism leads naturally to new,
improved rules of quantization (the quantum equation of motion
(\ref{neweqofmot}) remains unchanged though). We proposed the new,
improved rules of quantization:

\bea \varphi^a&\longrightarrow& \widehat{\varphi}^a=\varphi^a\nonumber\\
\pi_a&\longrightarrow& \widehat\pi_a= -i\frac{\p}{\p \varphi^a}
\label{improvedqrules}\eea

\section{Improved quantum fields}

For the Dirac field the improved rules of quantization reproduce
the same quantum theory as the old one (with the important
difference that now the classical theory is consistent). For the
Klein-Gordon field the change of rules of quantization gives rise
to minor but important differences.

The Lagrangian in Eq. (\ref{LagrangianS}) yields  the momenta

\be \pi = \dirac{\p}u\ee and the Hamiltonian

\be \H_S = \frac12(\pi^2 + u^2) \label{improvedHS}\ee Now, the
rules of quantization (\ref{improvedqrules}) yield the quantum
operator

\be\widehat{\H_S} = \frac12(-\frac{\p^2}{\p u^2} + u^2)
\label{improvedqHS}\ee

The quantum covariant Hamiltonian has been altered in that the
numerical factor multiplying the first term in the operator has
changed. The 4 (which would become $D$ for a $D$-dimensional
space-time) has disappeared.

That such a numerical factor (a $D$) appeared in the quantum
operator for the scalar field has already been noted in the
literature as an unsatisfactory feature of the previous formalism
\cite{[Kanatchikov2]}. That it does not appear when the improved
rules of quantization are used is another factor in their favor.

\section{Conclusions and other comments}

After a detailed discussion of the finite-dimensional formalism in
field theory we have applied the formulation proposed by M.
Navarro \cite{[towardfqft]} (and independently by I.V. Kanatchikov
\cite{[Kanatchikov]}) to the scalar field and the Dirac field. In
doing so several unsatisfactory aspects of the classical and
quantum formalism have become apparent. In order to overcome these
problems a modification of the classical as well as the quantum
formalism have been proposed.

In summary, our proposal goes as follows:

\noindent{\bf Classical finite-dimensional canonical formalism}
\v2mm \noindent Lagrangian: $\L =
\L(\varphi^a,\p_\mu\varphi^a)$\hfill\newline \noindent Momenta:
$\pi_a =
\gamma_\mu\frac{\p\>\L}{\p\>(\p_\mu\varphi^a)}$\hfill\newline
 \noindent
Hamiltonian: $\H (\pi_a,\varphi^a) = \pi_a \dirac{\p}\varphi^a -
\L$\hfill\newline
 \noindent Canonical Equations of motion:
\bea \dirac{\p}\varphi^a &=& \frac{\p \H}{\p\>\pi_a}\nonumber\\
 \dirac{\p}\pi_a&=& -\frac{\p
H}{\p\>\varphi^a} \label{improvedqrules2}\eea

\noindent{\bf Rules of quantization}

\bea \varphi^a&\longrightarrow& \widehat{\varphi}^a=\varphi^a\nonumber\\
\pi_a&\longrightarrow& \widehat\pi_a= -i\frac{\p}{\p \varphi^a}
\label{improvedqrules2}\eea

\noindent{\bf Quantum (Schr\"odinger like) equation of motion}

\be i\dirac{\p}\Psi = \widehat{\H}\Psi\ee

The new formalism solves three known problems of the previous
formalism:

\begin{itemize}

\item It restores the symmetry fields-momenta of the ordinary
canonical formalism.

\item It produces the right equations of motions for the Dirac
with the natural covariant Hamiltonian $\H = m\bar{v}v$.

\item It eliminates a numerical factor $D$ (the spacetime
dimension) from the quantum Hamiltonian operator of the scalar
field.

\end{itemize}

Further we have shown that the quantum equations of motion
correspond to fields with harmonic-oscillator-like internal
degrees of freedoms. It seems natural to predict that these extra
degrees of freedom will lead to a multi-particle interpretation of
the theory.

It has been shown that the formalism behaves consistently in
addition to being very elegant. The next step should be to produce
falsifiable predictions.

In this regard a satisfactory description of the electromagnetic
field should be produced. The difficulty here, of course, is how
to covariantly describe in a canonical way a theory that has
reduced degrees of freedom because of the gauge invariance
$A_\mu\longrightarrow A_\mu +\p_\mu\phi$. Of course, similar but
surely even more difficult problems are faced when applying the
formalism to gravity theories.

We hope to be able to report progress in these areas in the near
future.

 \v1mm

\noindent{\bf Acknowledgements.} The author is grateful to L.
Alvarado for her support, to M.M. de Fez Laso for the lending of
some very useful books, to M. Calixto for helping with submitting
the paper to the archives and to MiKTeX and WinEdt for their
magnificent word-processing software.

\end{document}